\documentclass[preprint]{elsarticle}
\usepackage{amsmath}
\usepackage{amssymb}
\usepackage{epsfig}
\usepackage{epstopdf}
\usepackage{graphicx}
\usepackage[latin5]{inputenc}
\usepackage{subfigure}
\usepackage{lineno}
\usepackage{hyperref}

\usepackage{color}

\numberwithin{equation}{section}

\begin{document}

\title{Unexpected parameter ranges of the 2009 A(H1N1) epidemic for Istanbul and the Netherlands}

\author[a]{Ali Demirci}
\ead{demircial@itu.edu.tr}

\author[a]{Ayse Peker Dobie}
\ead{pdobie@itu.edu.tr}

\author[b]{Ayse Humeyra Bilge }
\ead{ayse.bilge@khas.edu.tr}

\author[a]{Semra Ahmetolan}
\ead{ahmetola@itu.edu.tr}

\address[a]{Department of Mathematics, Faculty of Science and Letters, Istanbul Technical University, Istanbul, Turkey}
\address[b]{Department of Industrial Engineering, Faculty of Engineering and Natural Sciences,  Kadir Has  University, Istanbul, Turkey}

\begin{abstract}
The data of  the 2009 A(H1N1) epidemic in Istanbul, Turkey is unique in terms of the collected data, which include not only the hospitalization but also the fatality information recorded during the pandemic. The analysis of this data  displayed  an unexpected time shift  between the hospital referrals and fatalities. This time shift, which does not conform to the SIR and SEIR models, was explained by multi-stage SIR and SEIR models \cite{Dobie}. In this study we prove that the delay for these models is half of the infectious period within a quadratic approximation, and we determine the epidemic parameters  $R_0$ (basic reproduction number), $T$ (mean duration of the epidemic) and $I_0$ (initial number of infected individuals) of the 2009 A(H1N1) Istanbul and Netherlands epidemics.
These epidemic parameters were estimated by comparing the normalized cumulative fatality data with the solutions of the SIR model.  Two different error criteria, the $L_2$ norms of the error over the whole observation period and over the initial portion of the data, were used in order to obtain the best-fitting models. It was observed that, with respect to both criteria, the parameters of  "good" models  were agglomerated along a line in the $T$-$R_0$ plane, instead of being scattered uniformly around a "best"  model.
As this fact indicates the existence of a nearly invariant quantity, interval estimates for the parameters were given.
As the initial phase of the epidemics were less influenced by the effects of medical interventions, the error norm based on the initial portion of the data was preferred. The minimum error values for the Istanbul data correspond to the parameter ranges $4.2-5.4$, $13-15$ and $10^{.-7}-10^{-8.8}$ for  $R_0$, $T$ and  $I_0$, respectively. However, these parameter ranges are well out of the range for  the usual influenza epidemic parameter values. To confirm our observations on the Istanbul data,
the same error criteria were also used   for the  2009 A(H1N1) epidemic for the Netherlands, which has  a similar population density as in Istanbul. The minimum error values for the Netherlands data led   to the parameter ranges $2.9-4.2$, $3-5$ and $10^{-7}-10^{-9}$ for  $R_0$, T and  $I_0$, respectively. As in the Istanbul case, the parameter ranges do not match  the usual influenza epidemic parameter values.

\end{abstract}



\maketitle

\section{Introduction}
The 2009 A(H1N1) pandemic has been the subject of many studies.  Some of these works were limited to a specific country, such as Turkey \cite{Erg1}, Denmark \cite{Den}, Canada \cite{Can}, Morocco \cite{Mor} and Iran \cite{Ira} to report the spread of the epidemic. Some works were based on the comparison of the characteristic of epidemic on transnational basis \cite{Sur1},\cite{Wor1},\cite{Wor2},\cite{Bo},\cite{Sen}. In some other works,   the concentration  was  on the transmission dynamics,  the estimation of ``basic reproduction number", ``incubation period", ``generation time", and "serial interval" \cite{Bo},\cite{Sen},\cite{Tom},\cite{Whi},\cite{Fra},\cite{Muno},\cite{Sil},\cite{Bilge1},\cite{Bilge2}.

The original mathematical  model for the spread of epidemics is the integral equation model presented by  Kermack-McKendrick in 1927 \cite{Ker}.  Under the choice of the kernel in the form of $e^{-\gamma t}$ \cite{Heth80}, this model leads to systems of ODE's called the Susceptible-Infectious-Removed (SIR) and Susceptible-Exposed-Infectious-Removed (SEIR) models. In the framework of the standard SIR and SEIR epidemic models, the peak of the number of infected individuals coincides with the inflection point of the number of removed individuals. On the other hand, in the context of an extensive survey on the 2009 A(H1N1) epidemic conducted by a group of physicians \cite{Erg1},
had features that were not in agreement with this fact.
In this study, information on (adult)   A(H1N1) patients referred to major  hospitals in Istanbul, Turkey, were collected. Data included the day of referral ($869$ patients)  and the date of fatality  ($46$ patients).

The survey based on this  data  displayed  an unexpected time shift, a lag of $8$ days,  between  the  peak of the number of referrals and the inflection point of cumulative fatalities \cite{Bilge1}. This feature was  not in agreement with the standard SIR and SEIR models as noted above and therefore, needed explanation.

Recently, we used  multi-stage SIR and SEIR models that led to epidemic curves with a time-shift \cite{Dobie}. In fact, we proved that, if $T$ is the mean infectious period,   for an $n$-stage model,  the distance between the maxima of each infectious stage is $T/n$  of the   mean infectious period, within a second order approximation.  In the present work, we prove  that, within the same approximation scheme,  the time shift  between the peak of the number of infected individuals and the inflection point of the number of removed individuals is $T/2$.  Numerical simulations presented in \cite{Dobie} provided confirmations for both of these facts.

As far as we know, the Istanbul data is unique in terms of the collected data, which include not only the hospitalization but also the fatality information during the pandemic. This allows us to have available data on both the number of infected and the number of removed individuals of  epidemic models.
In general, cumulative fatalities are considered to be proportional to the number of removed individuals and the proportionality constant is called the death rate. But as this constant depends  on both the morbidity of the disease and  the quality of the health care system \cite{Bilge2}, one has to fit models to the normalized fatality data, as representatives of the normalized number of removed individuals.

As a first step, (normalized) the fatality data, as representative of the (normalized) number of removed individuals,  were analysed in order to determine the epidemic parameters basic reproduction number $R_0$, the reciprocal of the mean infectious period $T$, and the percentage of the people infected initially, $I_0=10^{-k}$.  Simulations were run over a wide range of parameters and the best-fitting models were selected. It has been proved in \cite{Bilge1} that there is a unique SIR model that fits the normalized data of removed individuals, thus one would expect that the best-fitting models would be aggregated around this value. However, the best-fitting  models, based on the $L_2$-norm of the error, are found to be scattered around a line, with $R_0$ in the range of $4.2-5.4$ and $T$ in the range of $13-15$ days.
This behavior, that is, the scattering of the data along a line, has been observed in \cite{Bilge3} in fitting the SIR model to the fatality data of the weekly ECDC surveillance reports \cite{ECDC} for $12$ European countries,  but for these countries the ranges are $1.2-2$ and $2-14$ for $R_0$ and $T$, respectively.

On the other hand, it is generally agreed  that $R_0$ is in the range of $1.2-2.5$ and the infectious period is around at most $7$ days. In the present work, we elaborate on the error criteria to explain this discrepancy.

To support our observations on the Istanbul data we also analyse the 2009 A(H1N1) epidemic data from the Netherlands, which like Istanbul,  has a high population density. As far as we know, no analysis has actually  been carried out for the Netherlands data (Table ~\ref{tab:my-table}), for which we used the weekly ECDC surveillance reports \cite{ECDC} due to the extensive incongruence in the data of ECDC annual reports (see Table 1 in \cite{Bilge2}).  The construction of a new data set includes the estimation of data by the use of an  interpolation method for the weeks 44-45 (2009) due to the absence of data and the week 46 (2009) due to the incongruence. Other weekly data are directly used from those weekly reports. Then the data were analysed  using the same error criteria to determine the  intervals for the epidemic parameters. We find $R_0$ around $2.9-4.2$ and $T$ in the range of $3-5$ days. These findings for the Netherlands are again  contrary to to the usual epidemic ranges  \cite{Bilge3}. To justify our arguments we compare the data for the Netherlands and Istanbul, and  as in the Istanbul case, it is also observed that values of  these three parameters can not be estimated consistently for the Netherlands as well.

In Section 2, we present the mathematical models for that standard and the  multi-stage SIR and SEIR models and we prove that the time shift between the peak of the infectious individuals and the inflection point of the number of removed individuals is half of the infectious period within a quadratic approximation.
In Section 3, we describe the Istanbul data, determine the related parameters for the SIR model using various error criteria and we show that parameters obtained by fitting models to the  initial phase of the epidemic data  give more realistic estimates. In the same section, we illustrate the time shift on Istanbul data. In Section 4, we also perform the same analysis for the 2009 A(H1N1) epidemic in the Netherlands.  Concluding remarks are presented in Section 5.

\section{Theoretical results and models}  

The  SIR and SEIR epidemic models without vital dynamics are defined by the following system of nonlinear  ordinary differential equations
\begin{eqnarray}
\label{eq:schemeP}
\begin{aligned}
SIR:~~~~S'&=-\beta SI\\
I'&=\beta SI-\gamma I\\
R'&=\gamma I
\end{aligned}\hskip1cm
\begin{aligned}
 SEIR:~~~~
S'&=-\beta SI\\
E'&=\beta SI-\epsilon E\\
I'&=\epsilon E-\gamma I\\
R'&=\gamma I
\end{aligned}
\end{eqnarray}
where the coefficients $\beta$, ${1}/{\gamma}$ and ${1}/{\epsilon}$   refer to the disease transmission rate, the mean  duration of infection period and the mean exposed period, respectively. Note that by the use of appropriate normalization one can choose  $S+I+R=1$ for the SIR model and  $S+E+I+R=1$ for the SEIR model.

The  multi-stage SIR and SEIR epidemic models corresponding to (2.1) are defined by the following systems of equations
\begin{eqnarray}
\label{eq:schemeP}
\begin{aligned}
SJR:~J&= I_0+\frac{\beta_1}{\beta_0} I_1+\cdots +\frac{\beta_N}{\beta_0} I_N\\
S'&=-\beta_0 SJ\\
I'_0&=\beta_0 SJ-\gamma_0 I_0\\
I'_i&=\gamma_{i-1} I_{i-1}-\gamma_{i} I_{i},\,\,for\,\,i=1,....N\\
R'&=\gamma_{N}I_N
 \end{aligned}
\hskip.5cm \begin{aligned}
SEJR:~J&=I_1+\frac{\beta_2}{\beta_1} I_2\cdots +\frac{\beta_N}{\beta_1} I_N\\
S'&=-\beta_1 SJ\\
E'&=\beta_1 SJ-\epsilon E\\
I'_1&=\epsilon E-\gamma_1 I_1\\
I'_i&=\gamma_{i-1} I_{i-1}-\gamma_{i} I_{i},\,\,for\,\,i=2,....N\\
R'&=\gamma_{N}I_N.
\end{aligned}
\end{eqnarray}
where $I_i$ is the density of individuals in the $i$th infectious stage with the  infectivity $\beta_i$, and  the infectious period  $1/\gamma_i$
satisfies the following conditions for the SIR model and SEIR models, respectively
$$
\frac{1}{\gamma_0}+\frac{1}{\gamma_1}+\cdots+\frac{1}{\gamma_N}=\frac{1}{\gamma}
$$
$$
\frac{1}{\gamma_1}+\frac{1}{\gamma_2}+\cdots+\frac{1}{\gamma_N}=\frac{1}{\gamma}.
$$

In one of our recent works we showed that  the systems defined by  (2.2)  explain the unexpected  time-shift between infectious and removed stages observed in the Istanbul data \cite{Dobie}. In that work, we also proved the following proposition \cite{Dobie}.

\textbf{Proposition 1:}  Let $t_i$ be the time where each substage  $I_i$ assumes its maximum and  $1/\gamma_{i}$ be the corresponding infectious period for $i=0,1,...N$. Then quadratic approximation provides that the successive difference  $t_{i}-t_{i-1}$ is   $1/\gamma_{i}$ and hence $t_{i}-t_{j}=\sum_{k=j+1}^{i} \frac {1}{\gamma_{k}}$.

In this paper we give the following proposition which proves that the   delay observed in the previous work is half of the infectious period.

\textbf{Proposition 2:} For the multi-stage SIR and SEIR epidemic models defined by the equations (2.2) where $\gamma_i=\gamma$ for all $i$, the delay between the peak of the total number of infectious individuals and the inflection point of the removed individuals is   half of the infection period within a quadratic approximation.

\vskip.2cm
\textbf{Proof :} Let the maximum of each infectious stage be $I_{im}$ for $i=0,1,\cdots,n$. For greater values of $n$, each infectious stage can be regarded as the shifted graph of the previous stage by $1/\gamma$ units to the right. Then one can express the following for $i=1,2,\cdots,n$
\begin{eqnarray}
\label{eq:schemeP}
I_i (t)=I_0(t-\frac{1}{\gamma})-(I_{0m}-I_{im}).
\end{eqnarray}
Considering $J(t)=I_0(t)+I_1(t)+\cdots +I_{n}(t)$ together with (2.3), one obtains
\begin{eqnarray}
\label{eq:schemeP}
J (t)=\sum_{j=0}^{n} I_0(t-\frac{j}{\gamma})-(n+1) I_{0m}+\sum_{j=1}^{n} I_{jm}
\end{eqnarray}
and consequently
\begin{eqnarray}
\label{eq:schemeP}
J' (t)=\sum_{j=0}^{n} I_0'(t-\frac{j}{\gamma}).
\end{eqnarray}
If the quadratic approximation of  Taylor series expansion of $I_0 (t)$ is used
at the point $t=t_0$ where it assumes its maximum value one obtains
\begin{eqnarray}
\label{eq:schemeP}
I_0 (t)=I_0 (t_0)+\frac{1}{2!} I_0'' (t_0)(t-t_0)^2
\end{eqnarray}
since $I_0' (t_0)=0$. Then differentiating (2.6) yields
\begin{eqnarray}
\label{eq:schemeP}
I_0' (t)=I_0'' (t_0)(t-t_0).
\end{eqnarray}
Substitution of (2.7) into the equation (2.5) gives
\begin{eqnarray}
\label{eq:schemeP}
J' (t)=I_0'' (t_0) \sum_{j=0}^{n}( t-\frac {j}{\gamma}-t_0).
\end{eqnarray}
Since the aim is to find the point where $J(t)$ assumes its maximum value, the equation in (2.8) is set as zero, and then considering the fact that $I_0''(t_0)\neq 0$ one obtains
\begin{eqnarray}
\label{eq:schemeP}
t=t_0+\frac {n}{2\gamma}.
\end{eqnarray}
Proof of the formula for the SEIR epidemic model is straightforward. $\Box$

\section{Determination of epidemic parameters  and the study of the time shift for the Istanbul  data}
\label{Standard epidemic models and epidemic models with multiple infectious stages}

In this section we study the determination of the  parameters of the SIR model  for the 2009 A(H1N1) epidemic in Istanbul, Turkey. The available data consist of hospital records collected at $6$ major state hospitals during the period of May 2009 and February 2010 \cite{Erg1}. The period covering the interval June 2009 and September 2009 is called ``the first wave" of the epidemic disease, during which  no  fatalities occurred and hospitalization rate was quite high.  The first wave is excluded from the data.  Thus, the work in consideration is based on the  period of $200$ days  covering the time interval September $1$, $2009$ to February  $28$,  2010. "The second wave" data include $869$ cases of hospital referrals after the exclusion of the first wave. Besides, the information includes the date of the referral to the hospital, the date when symptoms started according to patient's statement,  the date of discharge, which is the same as the referral date if the patient is not hospitalized, the date of transfer to the intensive care unit if applicable,  and the date of fatality.

In Figure 1, we display the raw data  for hospital referrals and for cumulative fatalities.

\begin{figure}[h!]
\centering
\subfigure[]{
\includegraphics[width=0.55\textwidth]{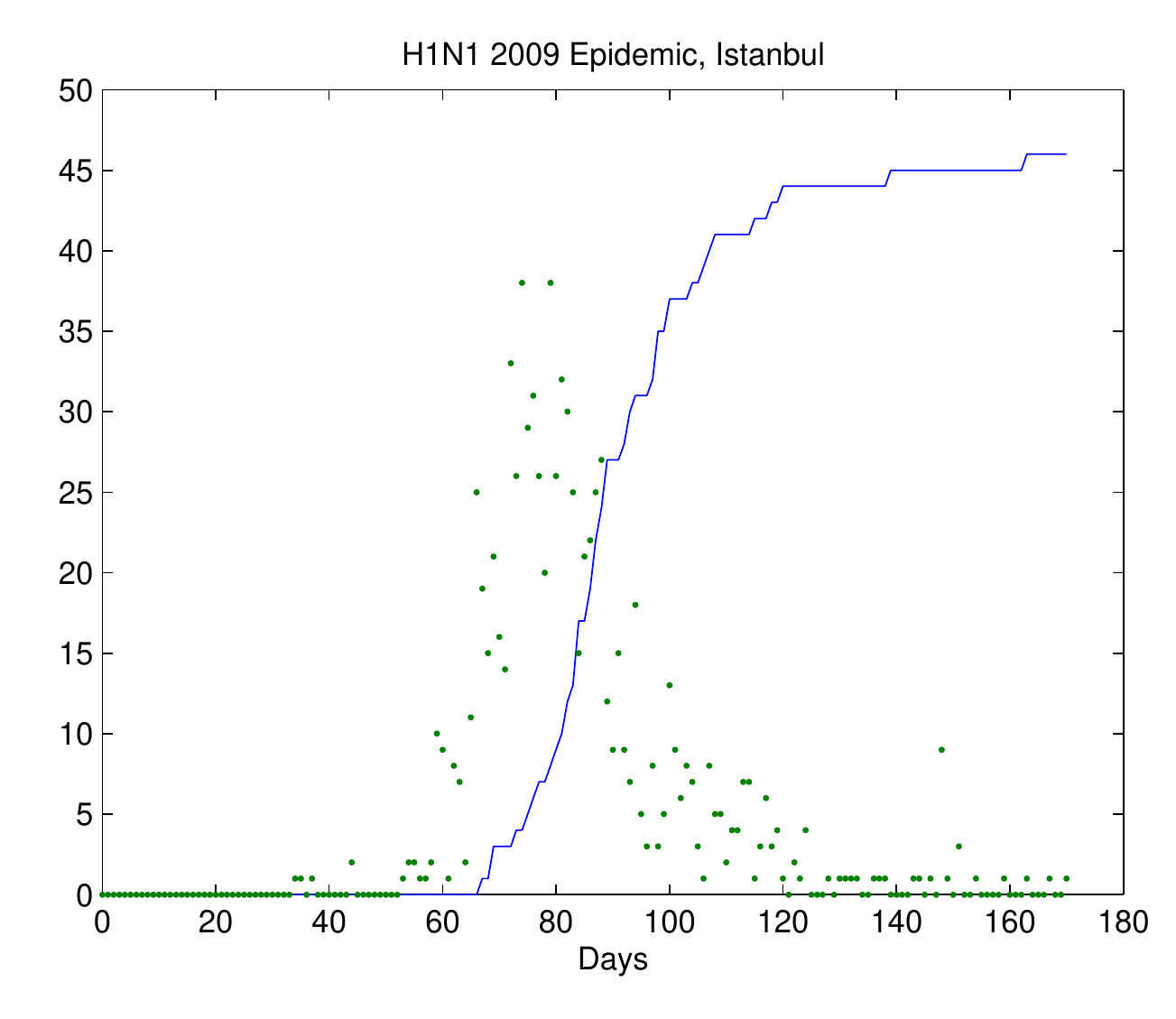}}
\subfigure[]{
\includegraphics[width=0.65\textwidth]{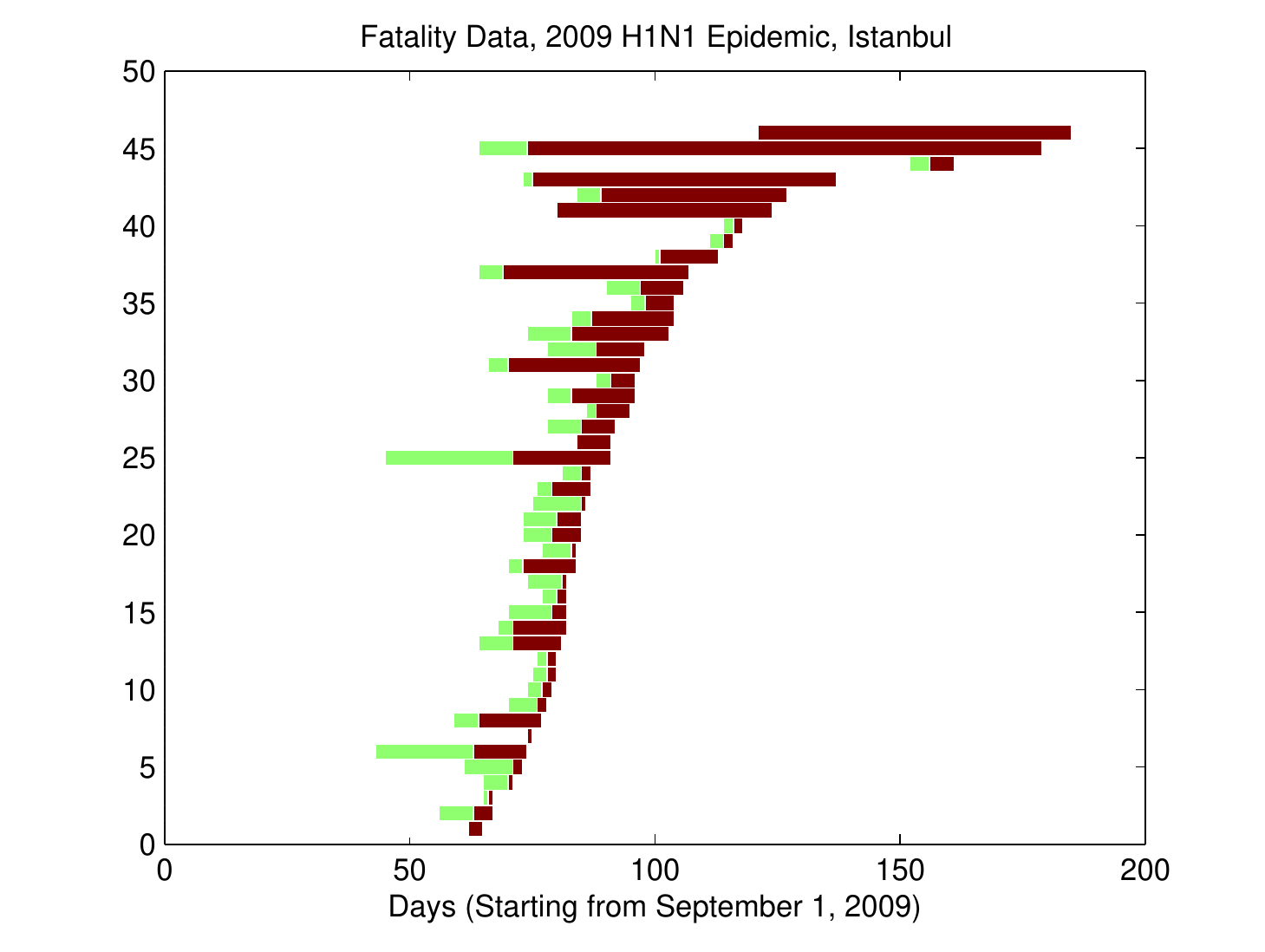}}
\caption{The daily number of referrals to hospitals and cumulative number of fatalities for the 2009 H1N1
epidemic in Istanbul, Turkey. The scatter in the number of daily referral to hospitals indicates that the
hospitalization rate varies during the epidemic. The asymmetry of the incidence curve is still observable.
The fatalities shown here are adjusted to at most 15 days after referral to the hospital (left panel). The duration
of symptoms prior to hospitalization (light color) and the duration of hospitalization prior to death (dark
color) for the fatalities in Istanbul, Turkey during due to 2009 H1N1 epidemic (right panel).}
\label{fig3}
\end{figure}

From this figure, we can see that hospital referrals are quite scattered.  On the other hand, we can clearly see that the peak of hospital  referrals lags behind the inflection point of cumulative fatalities.

Note that according to the SIR model, $I(t)$ is proportional to the rate of change in $R(t)$.
We quantify the lag in the Istanbul data by computing the correlation coefficients with lagged values and it is observed that maximum correlation is attained for a lag of $8$ days. By the proposition above, the infectious period is expected to be $15$ days, but this value falls outside reasonable bounds. We will show that epidemic parameters determined from the whole span of the normalized fatality data give $T$ values close to $15$ days, but the models based on the early phase of the data give more reasonable results.  This is in agreement with the observations in \cite{Bilge3}, where models based on the  early phase of the data were found to be better representatives of the spread of the epidemic.

The timing of the evolution of the epidemic is made by choosing September 1, 2009 as day 1, to day 200.  The values for $R(t)$ are the cumulative number of fatalities at that day. Although this counting leads to overestimates of the infectious period due to excessively long hospitalization periods, this is taken into account by giving less weight to errors towards the end of the epidemic. Thus, we use original data instead of correcting the long hospitalization periods. Besides, the cumulative number of fatalities are normalized, because the death rate is unknown. If we work only with the normalized $R(t)$ curve,  we know that the SEIR model fitting to the normalized data is not unique. Since we have also shown that the shapes of normalized $R(t)$ for SIR and SEIR models are practically indistinguishable, we use SIR model to determine the parameters by fitting models to the data and selecting the best models.

We start our analysis with the following parameter ranges: $1.5\leq R_0 \leq 8$ with  $0.1$ steps,    $2\leq T \leq 20$ with  $1$ steps and  $1\leq k\leq 10$ with $0.2$ steps where $I(0)=10^{-k}$ and we use the least squares error norm as the performance criteria
\begin{equation}
E(t)=\sqrt{\frac{(R_{s}/max(R_{s})-R_{r})^2}{R_{r}^{2}}}
\end{equation}
where $R_s$ and $R_r$ are the normalized fatality data obtained from SIR model and the real data, respectively.
But as mentioned above, extended hospitalization periods have to be taken into account. For this reason, we compute least squares errors for the beginning ($E_1:0-80$ days) and for the intermediate ($E_2:81-100$ days) periods separately. The error thresholds are chosen to be $E_1<0.04$ and $E_2<0.13$.

\begin{figure}[h!]
\centering
\includegraphics[scale=0.7]{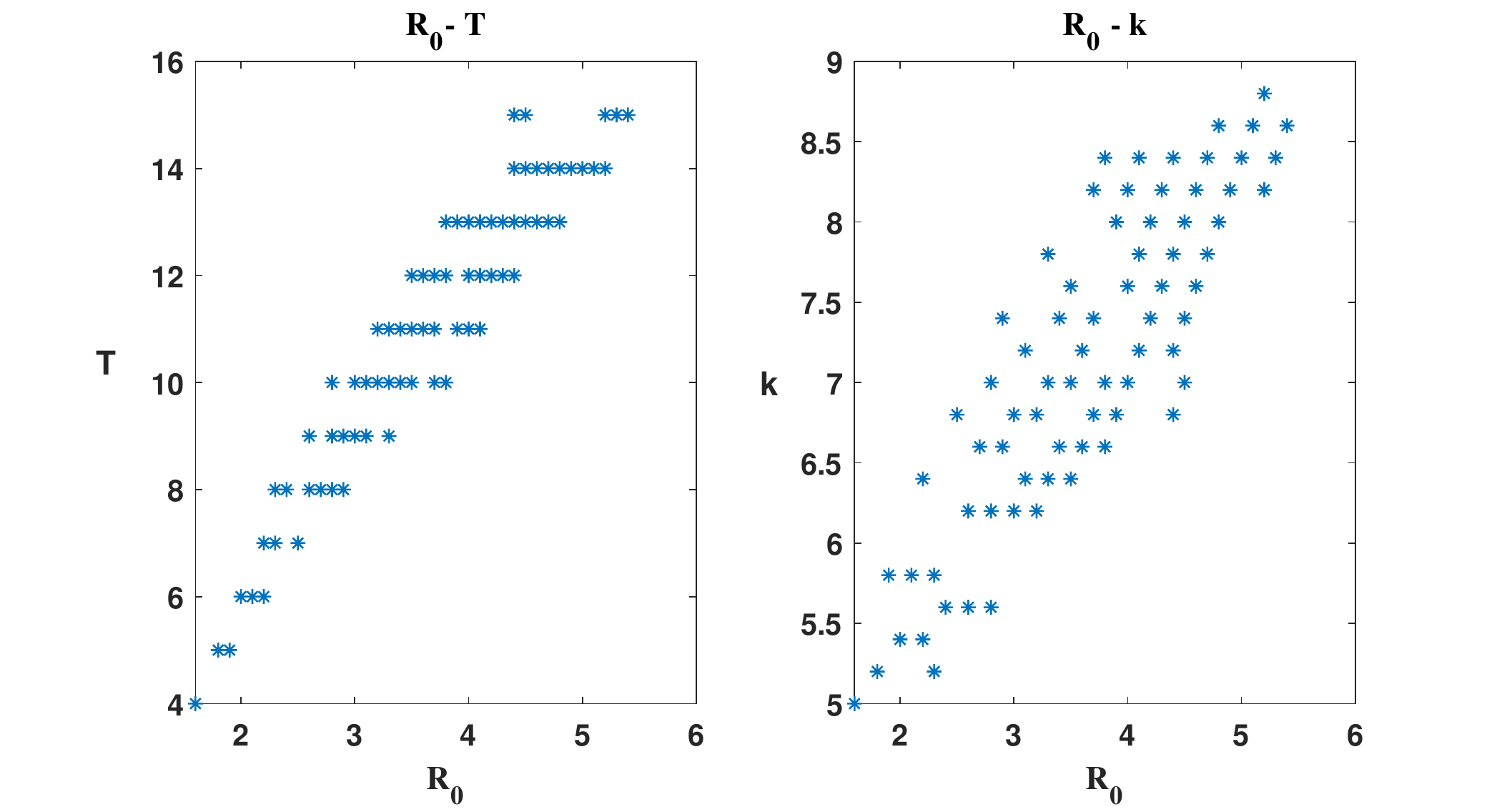}
\caption{Scatter plot of a) $R_0$ and $T$,  b) $R_0$ and $k$, for the Istanbul data.}
\end{figure}

The graphs of $R_0$ versus $T$ and $k$ are given in Fig. 2. We see in this figure that the best-fitting models range from low $R_0$, short $T$ for early starting epidemics to high $R_0$,  long $T$ for  late starting epidemics.  This shows that although a normalized SIR model fitting  curve is unique, in practice  there is some nearly invariant quantity.  In previous work \cite{Bilge2}, based on $R_0$ and $T$ plots, we have shown that this invariant quantity was the duration of the epidemic.  Here we give a surface in three dimensions (Fig. 3) to observe this fact.

\begin{figure}[h!]
\centering
\subfigure[]{
\includegraphics[width=0.5\textwidth]{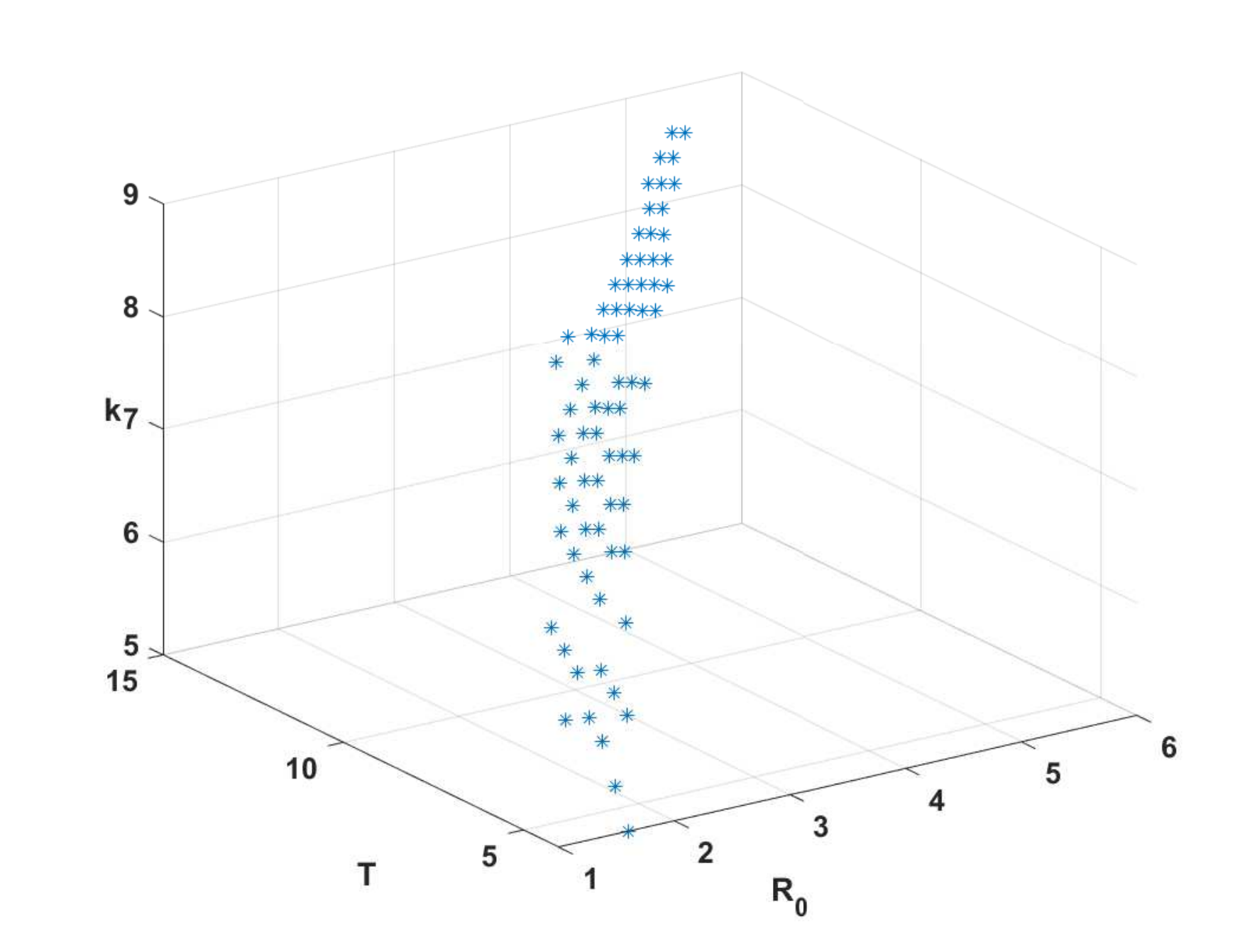}}
\subfigure[]{
\includegraphics[width=0.45\textwidth]{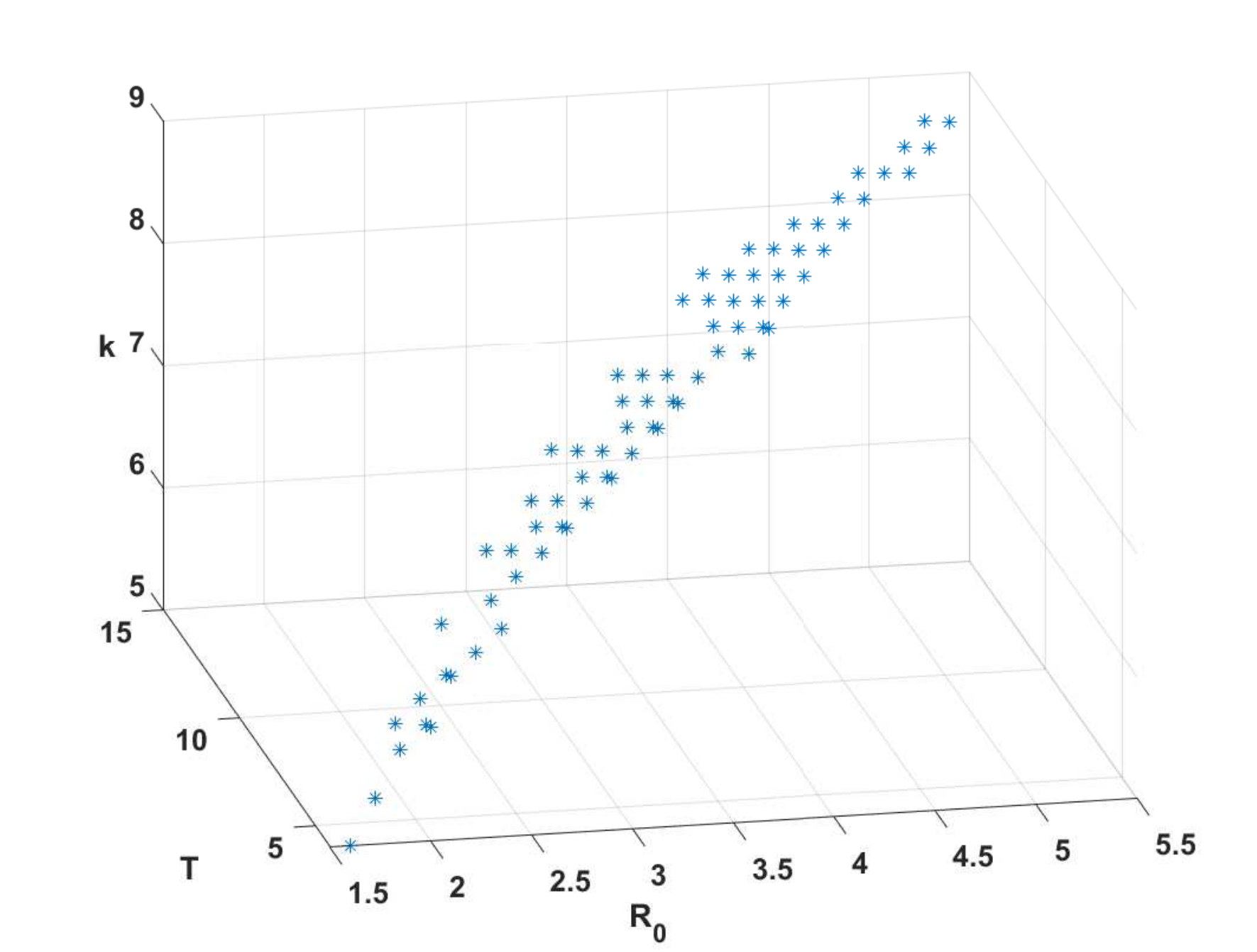}}
\caption{The scatter plot of $R_0$, $T$ and $k$ from different viewpoints for the Istanbul data. The scatter graph forms a surface (non-planar).}
\label{fig3}
\end{figure}

The variation of the parameters $R_0$, $T$ and $k$ subject to the specific error criteria  are plotted in Fig. 4.  $73$ different simulations for the parameter intervals  and for the error bounds are compared. As it can be seen in Fig. 4, the parameter intervals corresponding to the minimum error values are  $4.2-5.4$ for $R_0$,  $13-15$ for $T$ and  $10^{-7}-10^{-8.8}$ for  $I_0$. The interval of the infectious period $T$ conforms the lag value obtained from the Istanbul data and it is compatible with the result of the proposition that the lag is approximately half of the infectious period of the disease. Also from this figure, we see that high $R_0$ values  correspond to low total error while low $R_0$ values correspond to high errors  in the initial phases.

\begin{figure}[h!]
\centering
\includegraphics[scale=0.6]{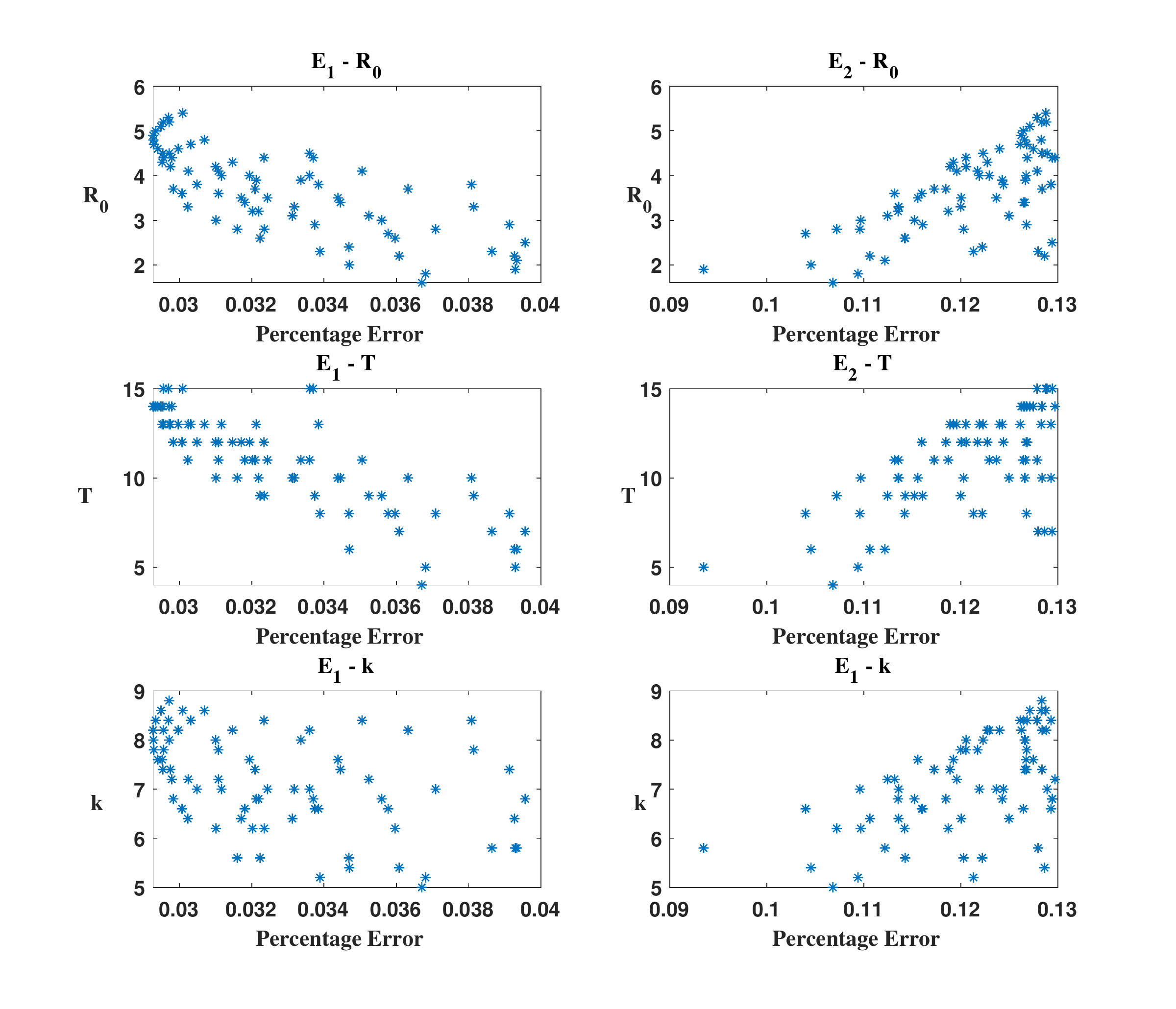}
\caption{Scatter plots of $R_0$, $T$ and $k$ versus $E_1$ and $E_2$ for the Istanbul data.}
\end{figure}

The values for $R_0$ obtained for  the  2009 A(H1N1) epidemic in other European countries vary between the boundaries of  $1.5-2.5$ \cite{Bilge2}. However, the range of $R_0$ for the Istanbul data falls far outside of this interval. In various studies the infectious period for influenza epidemic is about $T=3$ whereas  the value of $R_0$ stays in the interval $1.5-2.5$ \cite{Bilge3}. On the other hand, the values for  $R_0$ and  $T$ for the Istanbul data do not agree with the related values of the 2009 A(H1N1)  epidemic in most countries. Even though the values of the parameters do not fit the classical influenza parameter intervals, the range of the parameters for the  2009 A(H1N1) Istanbul epidemic are found exactly the same for each error criteria used. The results we obtain for the Istanbul data corresponding to the intervals for specific parameter values are compared with the ones for European countries in  Fig. 6 \cite{Bilge3}. The parameters which were found for the Czech Republic and Norway, are realistic; $R_0$ is about $1.5$ and $T<7$ days, but for Germany $R_0>2$ and $T>10$ days \cite{Bilge3}. Comparison of Fig. 2 and  Fig. 7 with  Fig. 5 shows that the scatter graph of $R_0$ for  Germany  is similar to those for Istanbul and the Netherlands.

\begin{figure}[h!]
\centering
\subfigure[]{
\includegraphics[width=0.30\textwidth]{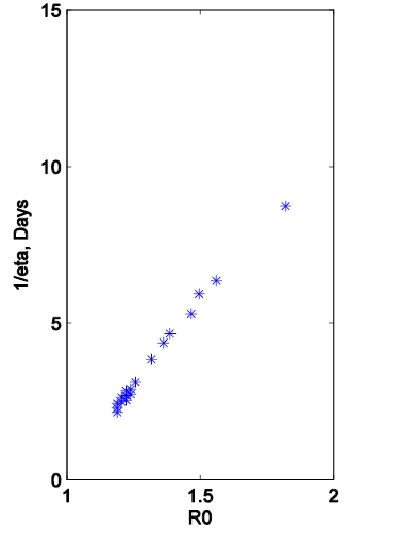}}
\subfigure[]{
\includegraphics[width=0.30\textwidth]{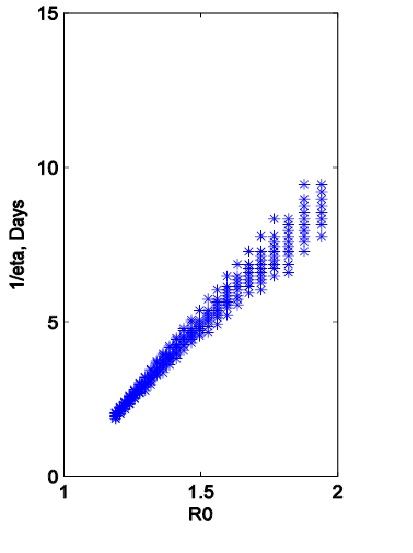}}
\subfigure[]{
\includegraphics[width=0.30\textwidth]{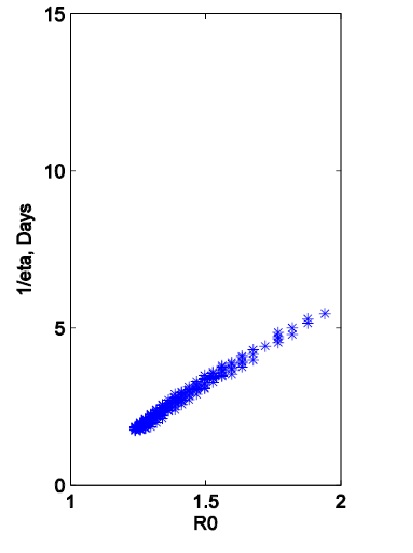}}
\caption{Scatter plot of $R_0$ and $T=1/\eta$ for the 2009 A(H1N1) epidemic in (a) Czech Republic, (b) Germany and (c) Norway.}
\label{fig3}
\end{figure}

\section{Determination of epidemic parameters  for the Netherlands data}
\label{Standard epidemic models and epidemic models with multiple infectious stages}

In a previous work \cite{Bilge3}, authors studied weekly fatality data for 13 European countries based on ECDC reports. The  Netherlands was considered for the analysis of vaccination coverage etc, but its data contained errors and it was excluded in the time domain analysis presented in \cite{Bilge3}.
In this section we analyse the Netherlands data, by first correcting for obvious errors in the data and then the study best models according to various error criteria.  In all cases $R_0$ turns out to be larger than the values reported in the literature.
\begin{table}[h!]
\begin{center}
\begin{tabular}{|c|c|c|c|c|c|c|l|}
\hline
Week  & N     & Week  & N     & Week  & N  & Week  & N     \\
\hline
36/09 & 1     & 46/09 & 24,49 & 03/10 & 59 & 13/10 & 63    \\
37/09 & 1     & 47/09 & 32    & 04/10 & 59 & 14/10 & 63     \\
38/09 & 5     & 48/09 & 37    & 05/10 & 60 & 15/10 & 63      \\
39/09 & 5     & 49/09 & 47    & 06/10 & 60 & 16/10 & 64     \\
40/09 & 5     & 50/09 & 54    & 07/10 & 60 & 17/10 & 64     \\
41/09 & 5     & 51/09 & 55    & 08/10 & 60 & 18/10 & 64      \\
42/09 & 5     & 52/09 & 56    & 09/10 & 60 & 19/10 & 64      \\
43/09 & 7     & 53/09 & 57    & 10/10 & 61 & 20/10 & 64      \\
44/09 & 11,54 & 01/10 & 57    & 11/10 & 62 & 21/10 & 65      \\
45/09 & 17,6  & 02/10 & 59    & 12/10 & 63 &       &              \\
\hline
\end{tabular}
\caption{Cumulative weekly fatalities ($N$) due to the A(H1N1) epidemic in the Netherlands from the ECDC weekly surveillance reports \cite{ECDC} between Week 36 (2009) and Week 21 (2010). This data set includes the estimation data by the use of an interpolation method for the weeks 44-45 (2009) due to the absence of data and the week 46 (2009) due to the incongruence.}
\label{tab:my-table}
\end{center}
\end{table}
Similar observations can be performed for the 2009 A(H1N1) the Netherlands data. The change in the total fatality number  in weeks according to the ECDC weekly reports  are given in Fig 6.
\begin{figure}[h!]
\centering
\includegraphics[scale=0.6]{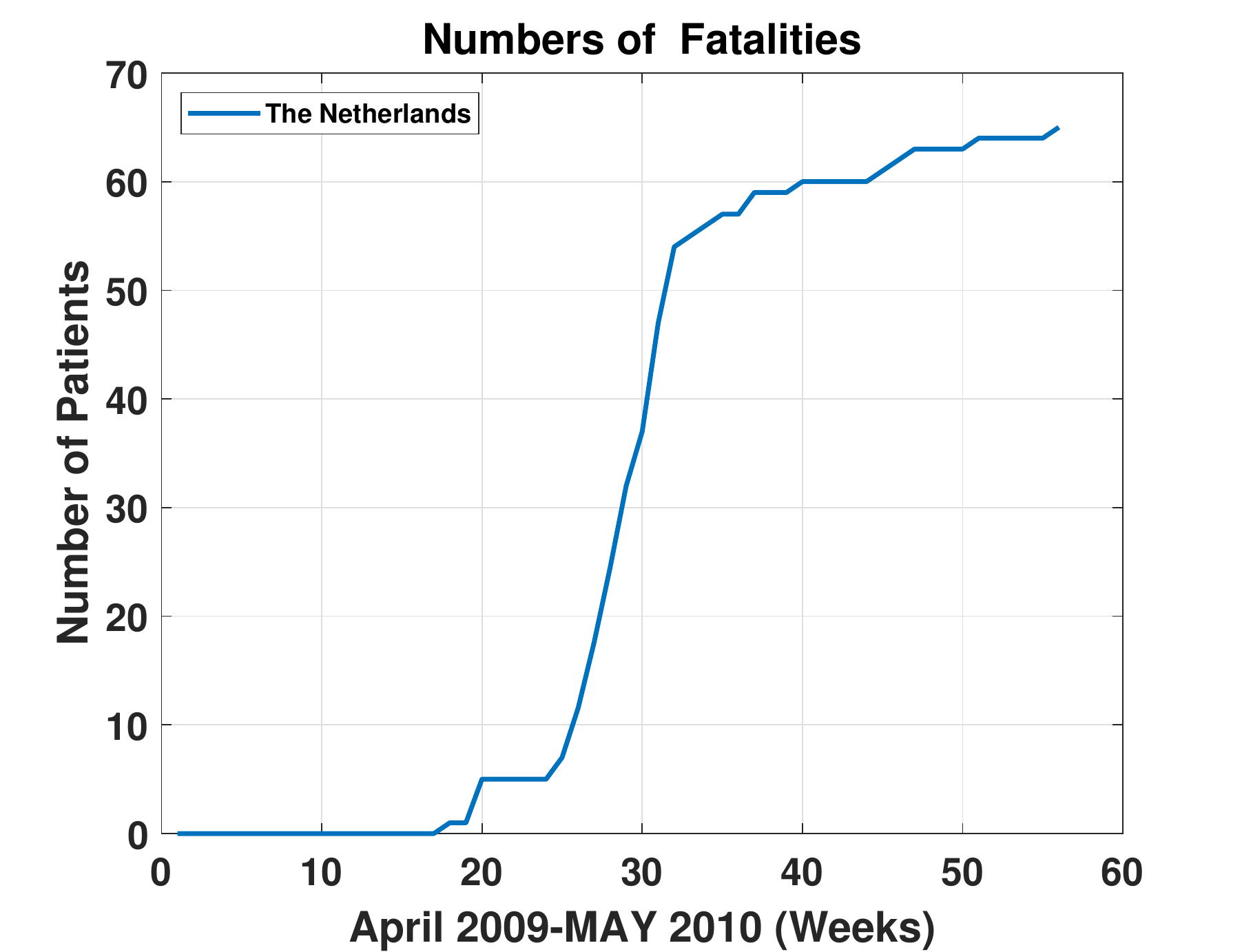}
\caption{Cumulative number of fatalities for the 2009 A(H1N1)
epidemic in the Netherlands.}
\end{figure}
Since there exists no hospitalization data for the Netherlands, the time shift phenomenon is not going to be discussed here. Furthermore, because of the reasons we stated for the Istanbul data, the normalized total fatality number for the Netherlands is going to be fitted to the  $R(t)$  curves which are  obtained by the numerical solutions of the SIR model. The periods used for the Netherlands data are taken $E_1: 0- 56$ weeks and $E_2: 0- 29$ weeks, respectively.  The error thresholds are chosen as $E_1<0.1$ and $E_2<0.19$.

For the Netherlands data, we use the following parameter ranges: $1.5\leq R_0 \leq 8$ with  $0.1$ steps,    $2\leq T \leq 20$ with  $1$ steps and  $1\leq k \leq 10$ with $0.2$ steps where $I(0)=10^{-k}$.

The graphs of $R_0$ versus $T$ and $k$ are given in Fig. 7. As in the Istanbul case, we see that the best-fitting models range from low $R_0$, short $T$ for early-starting epidemics to high $R_0$,  long $T$ for late-starting epidemics.  They form, in fact, a surface in $3$ dimensions, as shown in Fig 8.

\begin{figure}[h!]
\centering
\includegraphics[scale=0.65]{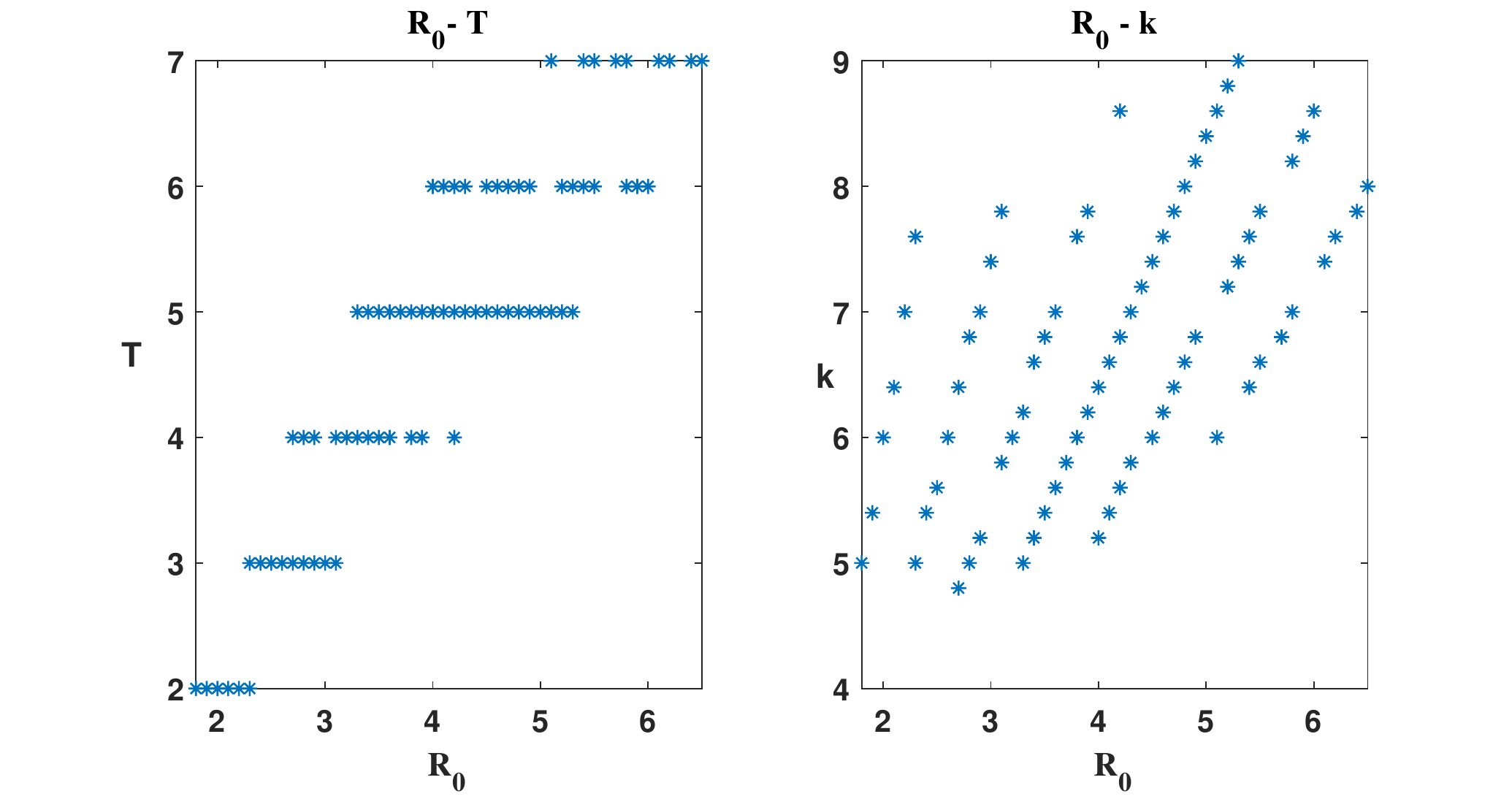}
\caption{Scatter plot of a) $R_0$ and $T$,  b) $R_0$ and $k$, for the Netherlands data.}
\end{figure}

\begin{figure}[h!]
\centering
\subfigure[]{
\includegraphics[width=0.5\textwidth]{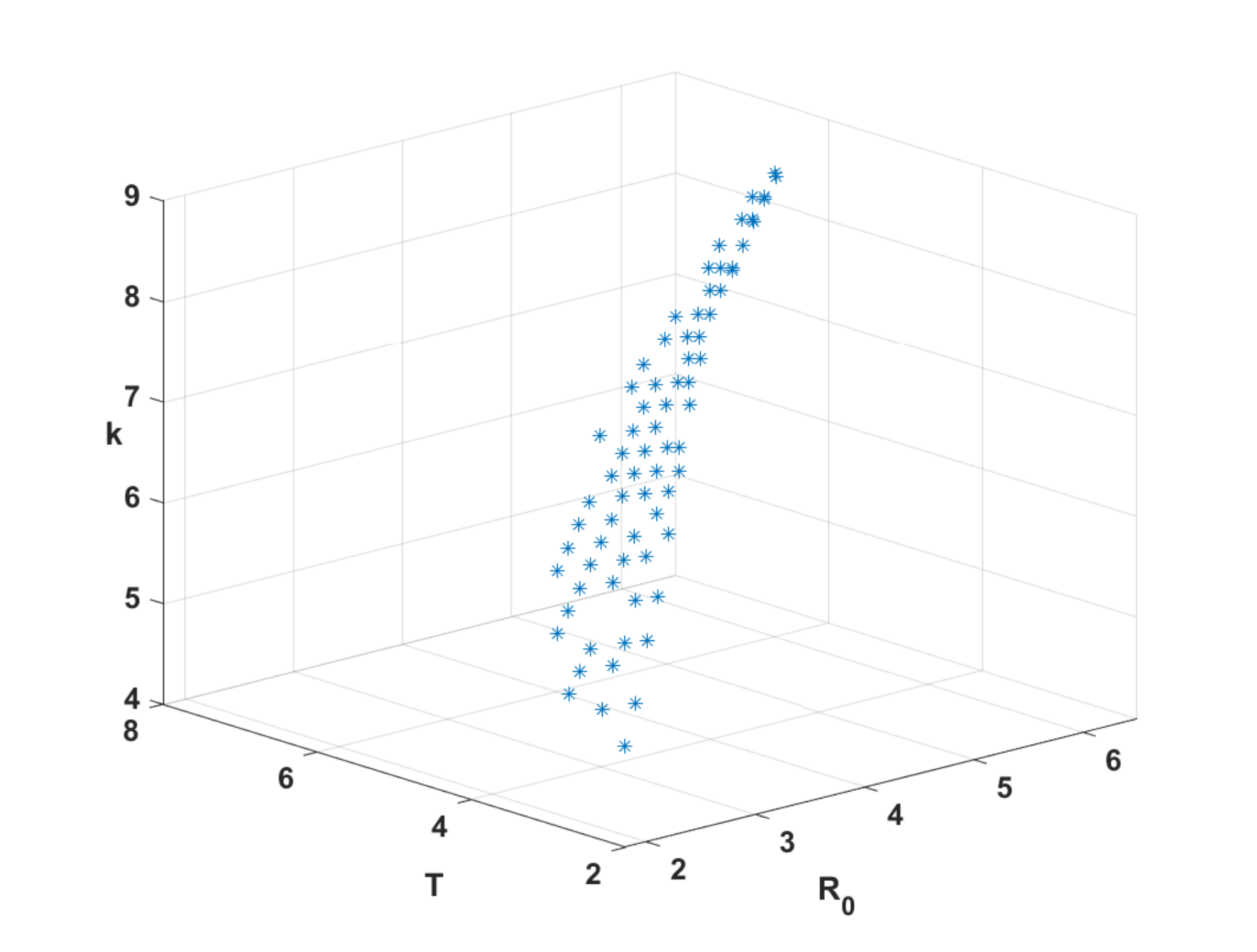}}
\subfigure[]{
\includegraphics[width=0.45\textwidth]{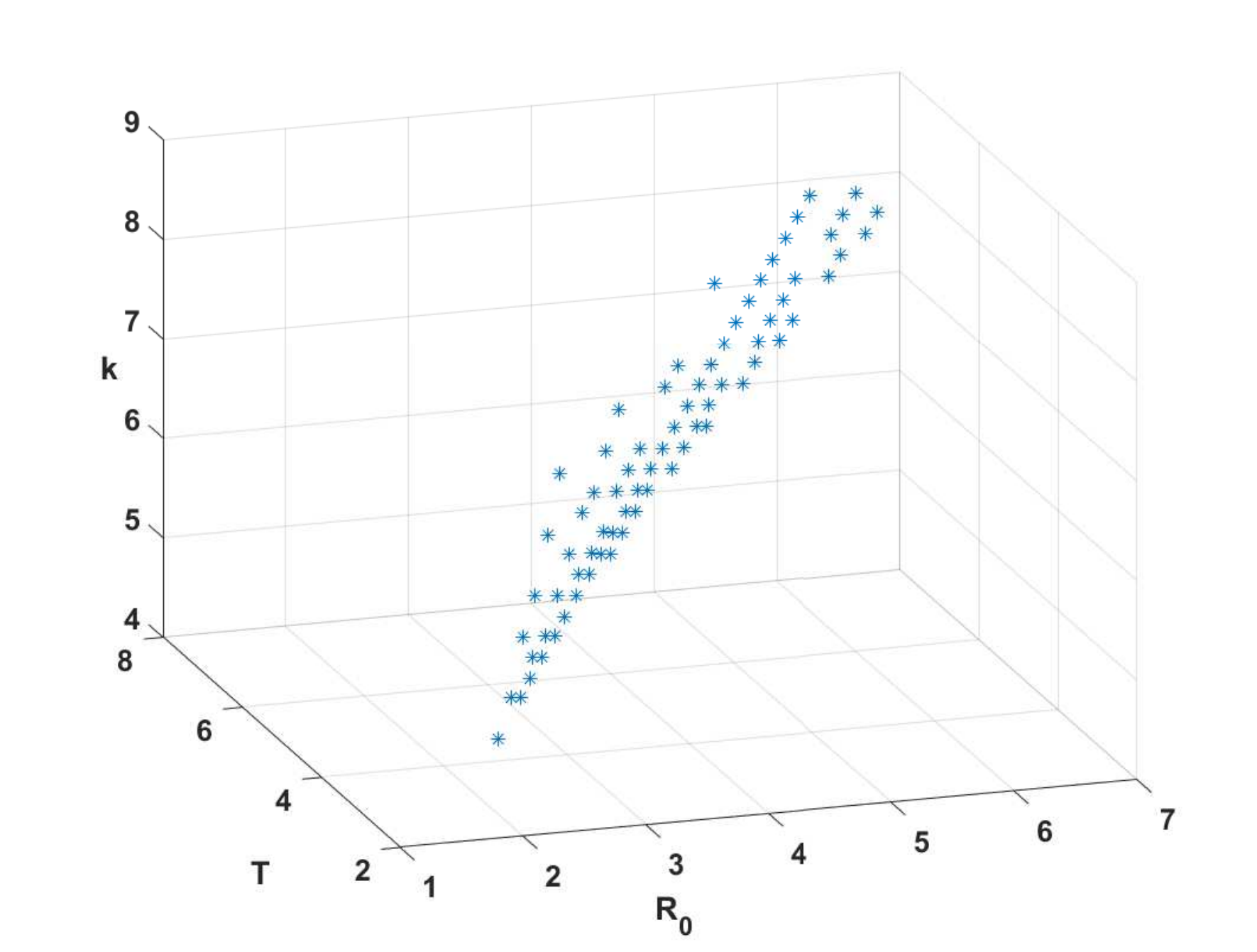}}
\caption{The scatter plot of $R_0$, $T$ and $k$ from different viewpoints for the Netherlands data. The scatter graph forms a surface (non-planar).}
\label{fig3}
\end{figure}

The variation of the parameters $R_0$, $T$ and $I_0$ subject to the specific error criteria for the Netherlands  are plotted in Fig. 9.  $73$ different simulations for the parameter intervals  and for the error bounds are compared. As it can be seen in Fig. 9, the parameter intervals corresponding to the minimum error values are $2.9-4.2$ for  $R_0$,  $3-5$ days for $T$ and  $10^{-7}-10^{-9}$ for $I_0$. Also, we see from the same figure that high $R_0$ values  correspond to low total error while low $R_0$ values correspond to high errors  in the initial phases as in the Istanbul case. Even though the infectious period interval for the Netherlands lies within reasonable bounds of influenza epidemics, the intervals observed for $R_0$ in  Istanbul or the Netherlands cases do not.
\begin{figure}[h!]
\centering
\includegraphics[scale=0.6]{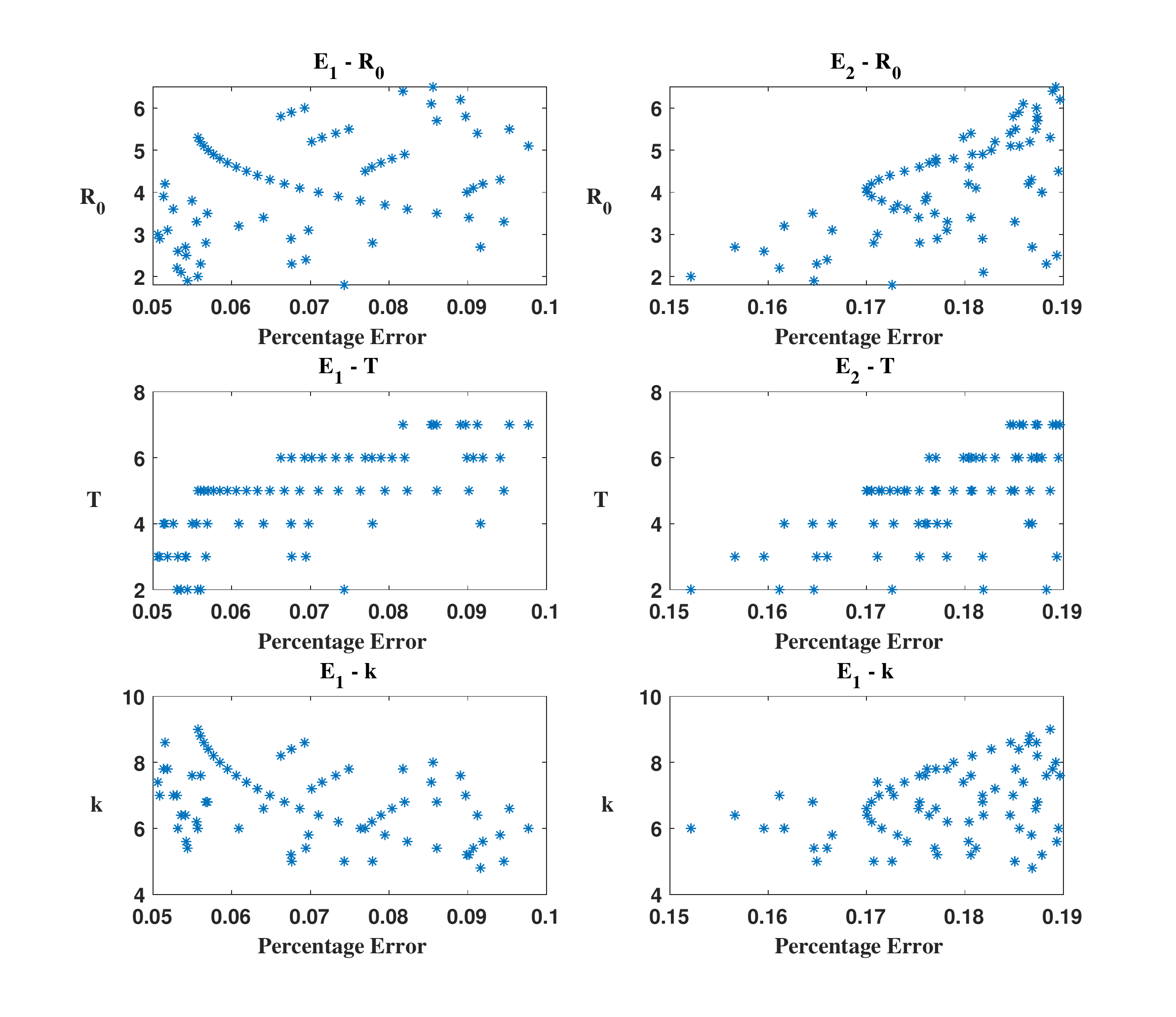}
\caption{Scatter plots of $R_0$, $T$ and $k$ versus $E_1$ and $E_2$ for the Netherlands data.}
\end{figure}

\section{Conclusion}
\label{Conclusion}
This article is concentrated on the data of the 2009 A(H1N1) epidemic in Istanbul and the Netherlands.  The survey based on the Istanbul data displayed an unexpected time shift between  hospital referrals and fatalities. This time shift was explained by the use of  multi-stage SIR and SEIR models \cite{Dobie}. In the present work, we prove analytically that the delay for these models is half of the infectious period.

Furthermore, in order to determine the epidemic parameters $R_0$, $T$ and $I_0$ of the 2009 A(H1N1) Istanbul epidemic we compare the normalized cumulative fatality data with the solutions of the SIR model. To obtain the best-fitting model two different error criteria are used.   The minimum error values for the Istanbul data correspond to the parameter ranges  $4.2-5.4$, $13-15$ and $10^{-7}-10^{-8.8}$ for  $R_0$, T and  $I_0$, respectively. However, these parameter ranges found through this analysis do not match  the usual influenza epidemic parameter values.  Therefore, we report that the 2009 A(H1N1)  Istanbul epidemic is an exception due to the results we obtain by using various error criteria.

In addition we also perform the same analysis for the 2009 A(H1N1) epidemic in the Netherlands which has a similar population density. To determine the parameter ranges we compare the normalized cumulative fatality numbers in  ECDC weekly reports with the solutions of the SIR model. To obtain the best-fitting model for the Netherlands data same two error criteria are used for different periods. As a result, the minimum error values for the Netherlands data correspond to the parameter ranges $2.9-4.2$, $3-5$ and $10^{-7}-10^{-9}$ for  $R_0$, T and  $I_0$, respectively.

Due to the existence of a nearly invariant quantity, values of these three parameters for  Istanbul and the Netherlands epidemics can not be estimated consistently by fitting a SIR model to such normalized epidemic data.

%
%
\bibliographystyle{elsarticle-num}
\biboptions{compress}

\end{document}